\documentclass[amsmath,floatfix,prb,reprint,superscriptaddress]{revtex4-1}
\usepackage{bm} % provides \bm{for boldmath}
\usepackage{hyperref} % provides url, href
\usepackage[T1]{fontenc}% if needed, must place before inputenc
\usepackage[utf8]{inputenc}
\usepackage{graphicx}
\usepackage{multirow}
\begin{document}
\author{I. S. S. de Oliveira}

\author{R. Longuinhos}
\email{raphael.lobato@dfi.ufla.br}
\affiliation{Departamento de F\'isica, Universidade Federal de Lavras, C. P. 3037, 37200-000, Lavras, MG, Brazil}

% Single-Crystal Monolayer Graphene
% Monolayer Graphene
% Bilayer Graphene
% Monolayer MoS2
% Designing electrical contacts to MoS2 monolayers: A computational study
\title{Effects of Oxygen Contamination on Monolayer GeSe: A computational study}

% exemplo de novo material 2D que oxida sem degradar?
% se é newly, discovered é redundante?
% the materials properties?
% Electronic-structure modifications induced by surface ssegregation
% Electronic structure modification induced by ion irradiation 
% results in severe, except in the top--Ge
\begin{abstract}
Natural oxidation is a common degradation mechanism of both mechanical and electronic properties for most of the new two-dimensional materials. From another perspective, controlled oxidation is an option to tune material properties, expanding possibilities for real-world applications. Understanding the electronic structure modifications induced by oxidation is highly desirable for new materials like monolayer GeSe, which is a new candidate for near-infrared photodetectors. By means of first-principles calculations, we study the influence of oxygen defects on the structure and electronic properties of the single layer GeSe. Our calculations show that the oxidation is an exothermic process, and it is nucleated in the germanium sites. The oxidation can cause severe local deformations on the monolayer GeSe structure and introduces a deep state in the bandgap or a shallow state near the conduction band edge. Furthermore, the oxidation increases the bandgap by up to 23\%, and may induce direct to indirect bandgap transitions. These results suggest that the natural or intentionally induced monolayer GeSe oxidation can be a source of new optoelectronic properties, adding another important building block to the two-dimensional layered materials.
\end{abstract}
\date{Submitted on \today}
\maketitle
% Titanium Oxide Nanosheets
% Graphene nanosheets
% Synthesis of graphene-based nanosheets via chemical
\section{Introduction}
The group IV monochalcogenide (GIVM) semiconductors family, based on germanium (Ge), tin (Sn) and the chalcogenides sulfur (S), sellenium (Se), and tellurium (Te), have promising properties for the development of a new generation of optoelectronic,\cite{SingleCrystalColloidal2010,NIRchottky2013,nanobeltsoxide2012,photoresponseGeSe2012} photovoltaic,\cite{tinandgermaniumsolarcells2011} and thermoelectric\cite{highefficientthermoelectric2015} devices. Furthermore, they are earth abundant\cite{earthabundance} and environmental friendly.\cite{tinandgermaniumsolarcells2011}
GIVM are pointed as an alternative for solar cells, as they are less toxic than other compounds used for photovoltaic applications, e.g.: Pb, Cd, Te. They can be synthesized by using cheaper routes like colloidal solutions, and have bandgaps in a suitable range for photovoltaic applications (1.1--1.5 eV).\cite{tinandgermaniumsolarcells2011} Besides, these compounds present interesting properties for the design of new thermoelectric devices, as anisotropic electronic and thermal conductivity.\cite{highefficientthermoelectric2015} GeSe is being envisioned for non-volatile memory applications,\cite{materialsnonvolatilememories2011,resistiveswitching2008} and GeSe nanosheet-based devices have shown promising properties for near-infrared photodetectors.\cite{NIRchottky2013}

% cleave exceptionally easily in the a-b plane -- prb
% Graphite can be easily cleaved 
The bulk GeSe crystal is a p-type semiconductor, described by an orthorhombic cell with the D$^{16}_{2h}$ space group and an unit cell with eight atoms, and two layers. Each layer has four three-fold coordinated atoms, with polar-covalent Ge--Se bonds, which form zig-zag chains along the minor crystal axis ({\bf b} [see Fig.~\ref{fig:pris}(a)]). The layers interact with each other by weak van der Waals interactions, and the crystal can be easily cleaved along the crystal axis perpendicular to the layers ({\bf c} [see Fig.~\ref{fig:pris}(b)]). Indeed, mechanical exfoliation of bulk GeSe has been reported to produce nanosheets of $\sim$ 57 nm thick.\cite{NIRchottky2013} The comparison of the bulk phase electronic dispersion along the in-plane and out-of-plane directions shows the 2D nature of its electronic properties.\cite{CoreExcitons1990}
%the onset of absorption for the GeS sheets in singlecrystalcolloidal2010
% Intramolecular Force Contrast and Dynamic Current-Distance Measurements at Room Temperature
Single-crystal nanosheets of GeSe were recently synthesized,\cite{SingleCrystalColloidal2010} revealing promising optoelectronic applications with an onset for optical absorption at about 1.075(5) eV~\cite{indirectabsorption1986,SingleCrystalColloidal2010} and indirect (direct) bandgap estimated to be 1.14 eV (1.21 eV).\cite{SingleCrystalColloidal2010} Similar values are reported on photoconductivity measurements at room temperature.\cite{photoconductivity1989}

% in low ambient conditions
% under ambient conditions
% thickness of 35 nm
% 35 nm thick nanobeltsoxide2012
% 5 nm thick lanthanum oxide thin films grown on Si(100) by
% during a two-minute water immersion
% High performance Si nanowire field-effect-transistors
% Ge/Si nanowire heterostructures as high-performance field-effect transistors
% Growth and synthesis of mono and few-layers transition
% Tribological characteristics of few-layer graphene
Usually, the manipulation and use of devices is carried under ambient conditions, and the samples may suffer oxidation depending on its reactivity. %In the monochalcogenides family, both bulk and monolayer GaSe is reported to readly oxide in ambient conditions.\cite{drapak2008native,tan2014electronic} 
GeSe nanobelts were observed to be coated by an 5 nm thick amorphous germanium oxide layer, which is dissolved after a few-minute water immersion.\cite{nanobeltsoxide2012} Although this loss of material does not affect the bulk crystal, it may have pronounced impact on few-layer GeSe devices. Germanium oxide is known to form on the surface of Ge-based materials. As an example, Ge nanowires readily oxidize upon air exposure, resulting in low-performance Ge nanowire-based field-effect-transistors.\cite{genanowireodixe2005} Thus, a better understanding of the oxidation effects on few-layer GeSe is highly desirable to support further developments on GeSe-based nanotechnology.

% the oxidation process
% We show that oxidation of phosphorene can lead to 
In this work we show that oxidation of monolayer GeSe is an exothermic process, and it can result on severe local geometry deformations at the oxidation sites. Our calculations indicate that oxidation occurs preferentially at germanium sites through Ge--O bonding. However, more complex defects involving oxygen bonding to both germanium and selenium are not ruled out. The resulting defects can introduce a deep state in the bandgap or a shallow state at the edge of the conducting band. Furthermore, the bandgap of the oxidized structures can increase by up to 23\% in comparison to the pristine monolayer, representing a blue-shift of the material absorption edge. In addition, the oxidation can promote a direct to indirect bandgap transition. These changes may be useful for customizing the material electronic properties.

\section{Methodology}
%The Brillouin Zone (BZ) integrations were performed within MP
We model the monolayer GeSe oxidation by using first-principles calculations, based on the density functional theory,\cite{hohenberg1964inhomogeneous,kohn1965self-consistent} as implemented in the \textsc{QUANTUM ESPRESSO} package.\cite{QE-2009} The Brillouin zone sampling are performed within Monkhorst-Pack scheme,\cite{monkhorst1976special} by using $2\times2\times1$ and $8\times8\times2$ $\Gamma$-centered grids for the supercell slab model ($3\times3\times1$) and bulk geometry relaxations, respectively, and a $12\times12\times1$ $\Gamma$-centered grid for the calculations of the electronic structure of the monolayer. We use a vacuum region of 14.0 \AA~in our slab model. For lattice and atomic position optimizations, convergence is achieved for forces and stresses lower than 0.1mRy/bohr and 50 MPa, respectively.
The exchange-correlation energy is evaluated with generalized gradient approximation (GGA), using Perdew-Burke-Ernzerhof parametrization.\cite{perdew1996generalized}
The valence electron-ion interactions are described using projector-augmented wave (PAW)\cite{blochl1994projector} potentials.
The plane-waves kinetic energy cutoff to describe the electronic wave functions (charge density) is set to 50 Ry (400 Ry).
To avoid spin contamination in the O$_2$ gas ground-state calculations, we perform spin polarized calculations, imposing the triplet configuration for the O$_2$ molecule.\cite{oxidizedCNT2000,oxidationCNT2003}

\section{Results and Discussion}

\begin{figure}
  \centering
  \includegraphics[width=0.48\textwidth]{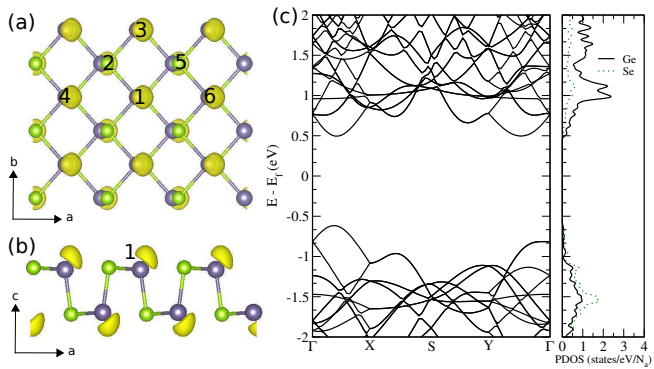}
  \caption{\label{fig:pris}(Color online) Pristine monolayer GeSe: (a) and (b) geometry and ELF (isovalue 0.93). (c) Electronic dispersion (left) and PDOS (right); the PDOS contribution of the Ge and Se atoms are in full black line and green-dotted line respectively.}
\end{figure}

%view from the $a\times b$ and $a\times c$ planes respectively
In Fig.~\ref{fig:pris}(a) and (b) we show the pristine monolayer GeSe geometry and the electron localization function (ELF)\cite{savin1997elf} plot (isovalue 0.92). Some of the germanium and selenium sites are labeled to further reference. As an example, the Ge(1) refers to the germanium at site (1) and Se(1) refers to the selenium at site (1), just below Ge(1). The ELF plot shows the Ge lone pairs, which are expected to be highly reactive. An ELF contour map along the Ge--Se bond (not shown) reveals its polar-covalent bond nature. % as expected from the Pauling negativity numbers (electronegativity) differences of the species [2.01 eV (Ge) \textit{.vs.} 2.55 eV (Se)].

%view from the $a\times b$ (a) and $a\times c$ (b) planes
% along the lines of high symmetry
% ELF só em b
% PDOS, passar para total e aproveitar melhor o espaço
% Passar fontes para mesmo tamanho dos numeros na estrutura

% ver bulk GeSebfsy.pwx-vcr.8121.lcc-ufla.in
% ver monolayer GeSeO_3x3x1-p.pwx-vcr.13578.vitor.lattice
The relaxed lattice parameters are $a=4.25$\AA~and $b=3.97$\AA, the crystal belongs to the C$_{2v}^7$ space group.
In comparison to the relaxed bulk GeSe lattice parameters, $a$ contracts by $+6.0$\% and $b$ expands by $+2.6$\%. Our calculated bulk lattice parameters values differ by $-3.2$\% and $+3.6$\% for $a$ and $b$, respectively, from the experimental values reported in Ref.~\onlinecite{refinement1978}, and by $-2.7$\% and $+4.2$\% for $a$ and $b$, respectively, from the experimental values reported in Ref.~\onlinecite{SingleCrystalColloidal2010}.
The monolayer lattice parameter modifications with respect to bulk values have also been observed on recent theoretical works.\cite{phophoreneanalogues2015}
% monolayer - bulk
%-3.2\% for $a$ and 3.6\% for $b$ in respect to bulk values~\cite{refinement1978} 
%-2.7\% for $a$ and 4.2\% for $b$ in respect to bulk values~\cite{SingleCrystalColloidal2010}.
% monolayer - bulk lidia
% a -4.3\%
% b 2\%
We observed the stretching of the Ge--Se bonds in the zig-zag direction by 2.3\%. The Ge--Se bonds along the $c$--direction decreases by 2.0\% and the height difference between the Ge and Se atoms at the zig-zag chain is 0.11 \AA~in the monolayer, representing a decrease of 57\% in comparison to the bulk value (0.26 \AA).
We calculate a direct bandgap of 1.11 eV, along the ${\bm{\Gamma}\bm{X}}$--direction on its Brillouin Zone [Fig.~\ref{fig:pris}(c) left] in agreement with calculations using similar methods.\cite{phophoreneanalogues2015} This value should be considered a lower bound for the experimental bandgap, which is unknown to our knowledge. From the projected density of states (PDOS) we see that the germanium states give the major contribution to the conducting bands [see Fig.~\ref{fig:pris}(c) right].

% oxidation of monolayer
% top--Ge -> GeSeO-01
% top Se -> GeSeO-02
%connecting two adjacent atoms of same species at the same zig-zag chain, through Ge--O--Ge (bridge-GeOGe; GeSe-05) and Se--O--Se (bridge-SeOSe; GeSe-03) bonds.
% Ge(1)--O--Se(5) bond (this configuration has the same relaxed geometry of the top--Ge),
% top--Ge  01& -0.780 \\
% bridge  02& -0.410 \\
% %side Ge 03&  0.611 \\
% bottom zig-zag (b-zz) 04 & -0.125 \\
% bridge Ge$_3$O 05& -0.994 \\

Interstitial oxygen can occupy several sites on monolayer GeSe and here we considered some of the high symmetry sites, as explained as follows. The oxygen is placed on top of Ge (top--Ge), on top of Se (top--Se), on the interstitial site along the Ge(1)--Se(1) bond [see Fig.~\ref{fig:pris}(b)], on the diagonal bridge site connecting two adjacent zig-zag chains through a Ge(1)--O--Se(5) bond, and connecting two adjacent atoms of same atomic specie on the same zig-zag chain, through Ge--O--Ge and Se--O--Se bonds.

To quantify which of these configurations are favorable, we adopt the following definition for chemisorption energy, per oxygen atom:\cite{oxygendefects2015}

\begin{equation}
  %E_\text{b} = \frac{1}{N_\text{O}}\left[E_\text{ox} - \left(E_\text{p} + \frac{N_\text{O}E_\text{O$_2$}}{2} \right)\right)
  E_\text{b} = E_\text{ox} - \left(E_\text{p} + \frac{E_\text{O$_2$}}{2} \right),
  \label{eq:Eb}
\end{equation}
where $E_\text{ox}$, $E_\text{p}$ and $E_\text{O$_2$}$ are the total energies of the oxidized monolayer GeSe, the pristine monolayer GeSe and the O$_2$ triplet molecule, respectively.
This definition results in negative $E_\text{b}$ for exothermic process and positive otherwise.
The binding energies per oxygen atom of the relaxed structures are summarized on Table~\ref{tab:eb}.

% table
% colocar multicollumn com initial and final
\begin{table}
 \centering
 \scriptsize
 \caption{\label{tab:eb} Binding energies per oxygen atom in the oxidized monolayer GeSe.}
 \begin{ruledtabular}
   \begin{tabular}{ccr}
     \multicolumn{2}{c}{Configurations} & $E_b$ (eV) \\
     Initial & Final & \\
     \colrule 
     top--Ge                            & \multirow{2}{*}{top--Ge}    &\multirow{2}{*}{ -0.780} \\
      Ge(1)--O--Se(5)                    &                         &        \\
      top--Se                            & Ge$_2$OSe bridge        & -0.410 \\
      Se--O--Se                          & not shown               &     0.611 \\
      interstitial Ge(1)-Se(1)           & bottom--zig-zag (b-zz)  & -0.125 \\
      Ge--O--Ge                          & Ge$_2$OGe bridge        & -0.994 \\
   \end{tabular}
 \end{ruledtabular}
\end{table}

% energy levels,
% defect states
% não defect bands
In general, the monolayer GeSe oxidation leads to minor changes on the lattice parameters.
%(expansion of $a$ by less than 4\% and contraction of $b$ by less than 2\%).
The major changes are on stretching, compression and breaking of Ge--Se bonds nearby the oxidation sites, which results in an defect state within the bandgap. We did not find magnetism induced by the oxygen defects in our calculations.

\begin{figure}
  \centering
  \includegraphics[width=0.48\textwidth]{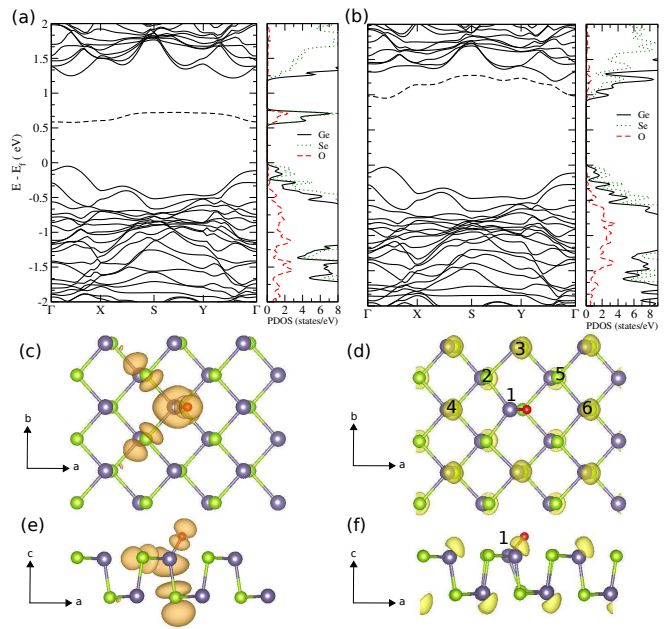}
  \caption{\label{fig:o101}(Color online) Top--Ge and constrained top--Ge. Electronic dispersion and PDOS for (a) constrained top--Ge and (b) relaxed top--Ge; the PDOS contribution of the Ge, Se, and O atoms are in full black line, green-dotted line and red-dashed line. (c) and (e) Geometry and LDOS (isovalue 0.002) related to the deep defect state in the constrained top--Ge bandgap. (d) and (f) Geometry and ELF (isovalue 0.92) of top--Ge.} %Geometries are view from $a\times b$ and $a\times c$ planes
\end{figure}

In the simulations with the oxygen starting on the diagonal bridge Ge(1)--Se(5), the Se(5)--O bond breaks remaining only the Ge--O bond, by using the electrons from the Ge(1) lone pair. In the resulting relaxed structure (top--Ge), shown in Fig.~\ref{fig:o101}(d) and (f), the Ge(1) has a tetrahedral-like coordination, resembling that of germanium atoms in bulk phase, and this coordination is expected to be energetically favorable.
%The structure of top--Ge presents mirror symmetry with respect to a plane containing the Ge(1)--O bond.
In comparison to pristine monolayer GeSe, the Ge(1)--Se(2) bond contracts by 6.0\%, the Se(2)--Ge(3) bond expands by 3.0\%, the Ge(1)--Se(5) distance increases by 13.1\%, the Ge(1)--Ge(4) distance decreases by 7.2\%, the Ge(1)--Ge(6) distance increases by 10.0\%, and the Ge(1)--Se(1) bond contracts by 2.0\%. The top--Ge structure is the second most energetically favorable, 21.0\% less exothermic than the most favorable (see Table~\ref{tab:eb}).

To gain insight on the electronic structure modifications in the monolayer GeSe due to oxidation, as observed in the top--Ge configuration, we considered a single oxygen atom free to move over a frozen monolayer GeSe.
The oxygen diffuses on the surface and binds to the Ge(1) lone pairs, as shown in Fig.~\ref{fig:o101}(c) and (d) and we refer to the resulting geometry as constrained top--Ge.
The comparison of the electronic dispersion of the constrained top--Ge [Fig.~\ref{fig:o101}(a) left] with the pristine GeSe [Fig.~\ref{fig:pris}(c) left] shows the influence of the Ge(1)--O bond only, as they are related to the same monolayer GeSe geometry.
Even in the absence of structural defects on the monolayer GeSe, a deep flat band is noticed within the bandgap of the constrained top--Ge [dashed line in Fig.~\ref{fig:o101}(a) left]. This band is composed by oxygen, germanium and selenium states, as shown by the PDOS in Fig.~\ref{fig:o101}(a) right. The related local density of states (LDOS), shown in Fig.~\ref{fig:o101}(c) and (e), provides further insight on the localized nature of this band.
The relaxation of the constrained top--Ge structure results in the top--Ge geometry, and the deep defect state becomes a shallow state, near the conducting band edge [dashed line in Fig.~\ref{fig:o101}(b) left]. In comparison to the pristine monolayer, the bandgap increases by 10\% and becomes indirect. This blue shift of the optical absorption edge comes through exothermic structural changes and should be irreversible unless energy is given to the system. In Ref.~\onlinecite{hu2015gese}, it is shown that direct to indirect bandgap transitions can also be achieved by controlling the strain along the system.

%GeSeO-01

\begin{figure}
  \centering
  \includegraphics[width=0.48\textwidth]{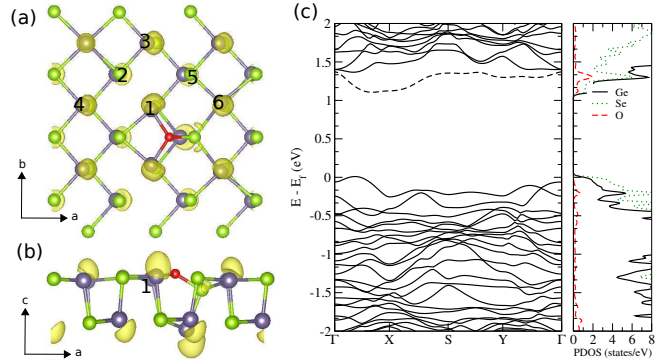}
  \caption{\label{fig:o102}(Color online) Ge$_2$OSe bridge. (a) and (b) Geometry and ELF (isovalue 0.92). (c) Electronic dispersion and PDOS; the PDOS contribution of the Ge, Se and O atoms in full black line, green-dotted line and red-dashed line, respectively.}
\end{figure}

% nearby atoms
A possible relaxation route of the initial top--Se configuration is through the Se(5)--O bond stretching, which allows the oxygen to bind with the two germanium atoms of the adjacent zig-zag chain, forming a Ge$_2$OSe bridge, as shown in Fig.~\ref{fig:o102}(a) and (b). %The structure has the same mirror plane, as the top--Ge configuration.
The lone pairs of the two germanium atoms bounded to the oxygen are slightly affected [Fig.~\ref{fig:o102}(a) and (b)]. The charge to form the Ge--O bond comes from the Ge(1)--Se(2) bond, which breaks (stretches by 15\%).
This configuration is about 59\% less energetically favorable than the most favorable configuration (Table~\ref{tab:eb}).
The oxygen defect introduces a shallow defect state near the conducting band edge [dashed line in Fig.~\ref{fig:o102}(a) left], and it is essentially composed by the oxygen and nearby germanium and selenium atoms, as seen from the PDOS [red-dashed line in Fig.~\ref{fig:o102}(c) right]. The absorption edge of this configuration is also blue-shifted, as the bandgap increases by $\sim$ 23\%.
 
%GeSeO-02

The oxidation of monolayer GeSe only through Se--O bond formation is unlikely. Indeed, the Se--O--Se configuration doest not lead to a favorable configuration (not shown), and experimental chemical composition measurements in GeSe nanosheets~\cite{NIRchottky2013} indicate only Ge--O bonds. Nevertheless, oxygen binding to both germanium and selenium (Ge--O--Se) has been observed in germanium-selenide glasses.\cite{geseglasses} 

\begin{figure}
  \centering
  \includegraphics[width=0.48\textwidth]{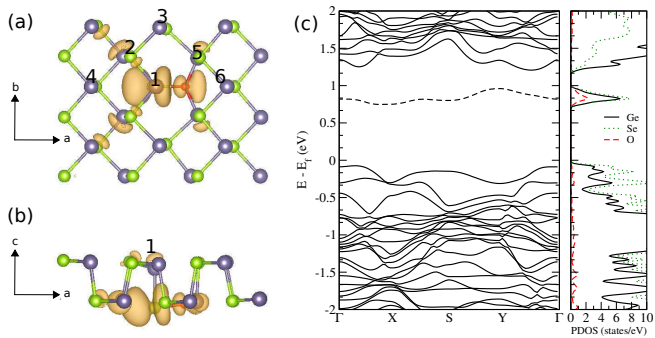}
  \caption{\label{fig:o104}(Color online) Bottom zig-zag (b-zz). (a) and (b) Geometry and LDOS (isovalue 0.002). (c) Electronic dispersion and PDOS; the PDOS contribution of the Ge, Se and O atoms are in full black line, green-dotted line and red-dashed line, respectively.}
\end{figure}

In the configuration with the initial interstitial Ge(1)--Se(1) oxygen defect, the oxygen migrates to the bottom zig-zag (b-zz) chain, breaking the Ge(5)--Se(1) bond to form Ge(5)--O and Se(1)--O bonds, as shown in Fig.~\ref{fig:o104}(a) and (b). The b-zz configuration is about 87\% less energetically favorable than the most favorable geometry (see Table~\ref{tab:eb}), but yet exothermic.
The comparison of the geometries shown in Fig.~\ref{fig:o102} and Fig.~\ref{fig:o104} indicates that it is more favorable for the oxygen to form a bridge between the two adjacent zig-zag chains than to enter in the Ge(5)--Se(1) bond of a single zig-zag chain.
The oxygen defect compress the Ge(1)--Se(2) bond by $\sim$3\%, stretches the Se(2)--Ge(3) bond by $\sim$2\%, and stretches the Ge(5)--Se(3) bond by $\sim$6\%. The Se(1) is pushed by the oxygen and the lateral distance (along a--direction) between Se(1) and Se(3) is $\sim$0.67~\AA.
These structural modifications introduce a deep defect state in the bandgap [dashed line in Fig~\ref{fig:o104}(c) left].
The bandgap increases by $\sim$ 15\% and it becomes indirect.
This defect state is essentially localized at the oxygen and nearby germanium and selenium atoms [Fig.~\ref{fig:o104}(a) and (b)].

%GeSeO-04

\begin{figure}
  \centering
  \includegraphics[width=0.48\textwidth]{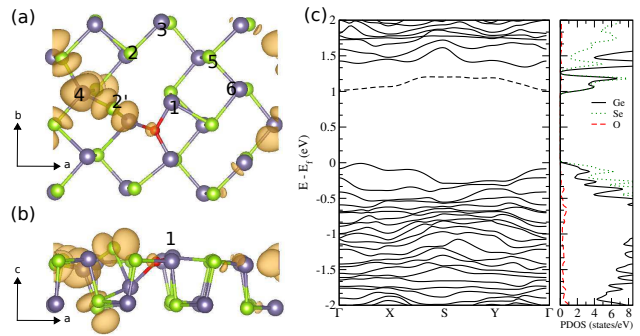}
  \caption{\label{fig:o105bgle}(Color online) Ge$_2$OGe bridge. (a) and (b) Geometry and LDOS (isovalue 0.002). (c) Electronic dispersion and PDOS; the PDOS contribution of the Ge, Se and O atoms are in in full black line, green-dotted line and red-dashed line, respectively.}
\end{figure}

% the most energetically favorable
% are energetically the most favorable
% most stable
% are energetically favorable
% the energetically most favorable configuration
In the energetically most favorable configuration, the oxygen connects two adjacent zig-zag chains through bonds with three Ge atoms (Ge$_2$OGe bridge). This configuration is achieved through a severe deformation of the monolayer GeSe. The Ge(1)--Se(2) and Ge(4)--Se(2) bonds break (stretched by 41\% and 39\%, respectively), the oxygen pulls up the Ge(2') bonding together, and the Se(2') moves to form the Ge(4)--Se(2') bond. In the resulting configuration, shown in Fig.~\ref{fig:o105bgle}(a) and (b), the Ge(4) assumes a tetrahedral-like coordination, and it is bounded to four Se atoms.
The oxygen defect introduces a shallow defect state near the conduction band edge [dashed line in Fig.~\ref{fig:o105bgle}(c) left]. The bandgap increases by $\sim$ 22\% and becomes indirect. Remarkably, this band has essentially no contribution from the oxygen atom, as can be seen from the PDOS [red-dashed line in Fig.~\ref{fig:o105bgle}(c) left] and are mainly localized around the Ge(4), as shown in the LDOS [Fig.~\ref{fig:o105bgle}(a) and (b)].

%GeSeO-05 

%The former examples should not sustain the argument that the Ge--O bond formation is always a favorable process. Indeed, in Fig.~\ref{fig:o103bg} we show a case where the oxygen is bound to a single germanium and the process is endothermic, thus very unlikely considering oxygen gas as source of oxygen. We note that the Ge--O bond is at the $a\times b$ plane, and it breaks the Ge--Se bond at the zig-zag chain, repealing the Se atoms which forms stretched Ge--Se bonds with the germanium atoms of the adjacent zig-zag chain. This defect introduces a deep defect band in the bandgap, which is related to states localized on the oxygen and nearby germanium and selenium atoms.

%GeSeO-03
%\begin{figure}
%  \centering
%  \includegraphics[width=0.48\textwidth]{geseo-03-bg.eps}
%  \caption{\label{fig:o103bg}(Color online) Electronic dispersion along high symmetry lines of the Brillouin Zone and PDOS of the bridge side Ge 03 configuration. In the PDOS, Ge states are in full black line, Se states are in green-dotted line and O states are in red-dashed line. The geometry is view from $a\times b$ and $a\times c$ planes.}
%
%\end{figure}

%The low dispersive lines within the bandgap may induce recombination mechanisms or trapping carriers, which result in lower performance on photo-based GeSe devices.

\section{Conclusion}
The oxidation process on monolayer GeSe is investigated using first-principles calculations. Our results indicate that the oxidation of the monolayer GeSe is an exothermic process, taking place preferentially at the germanium site. The introduction of oxygen on the monolayer GeSe can result in severe local geometry deformations, giving rise to a shallow or a deep defect state within the material bandgap. Furthermore, the results suggest that oxygen defects increase the monolayer GeSe bandgap, blue-shifting its absorption edge by up to 23\% and may induce direct to indirect bandgap transitions. Although the monolayer GeSe oxidation occurs naturally, if performed in a controlled way the oxidation process can possibly be a route for tunning the material electronic properties.

\begin{acknowledgements}
The authors acknowledge support from the Brazilian agencies CNPq, FAPEMIG and the computational time spent at LCC-UFLA. R. Longuinhos thanks to J. Ribeiro-Soares, for useful discussions on the manuscript.
\end{acknowledgements}
% bibliography

%The authors state that they were not aware of the work by Gomes L. \emph{et al.}, recently deposited in arXiv (http://arxiv.org/abs/1604.04092v1) . Although both works are similar, we consider they are simultaneous and independent productions.
\bibliography{GeSeoxide.bib}
\end{document}